\documentclass[prd,aps,showpacs,nofootinbib,nobibnotes,superscriptaddress]{revtex4}

\usepackage{epsfig}
\usepackage{amssymb}
\usepackage{bm}

\begin{document}

\title{The Spectrum of Gravitational Radiation from Primordial Turbulence}

\date{\today}
\author{Grigol Gogoberidze}
\email{gogober@geo.net.ge}
\affiliation{Centre for Plasma Astrophysics,
K.U.\ Leuven, Celestijnenlaan 200B, 3001 Leuven, Belgium}
\affiliation{National Abastumani
Astrophysical Observatory, 2A Kazbegi Ave, GE-0160 Tbilisi,
Georgia}
\author{Tina Kahniashvili}
\email{tk44@nyu.edu}
\affiliation{Center for Cosmology and Particle Physics, New York
University, 4 Washington Plaza, New York, NY, 10003 USA}
\affiliation{National Abastumani Astrophysical Observatory, 2A
Kazbegi Ave, GE-0160 Tbilisi, Georgia}
\author{Arthur Kosowsky}
\email{kosowsky@pitt.edu}
\affiliation{Department of Physics and
Astronomy, University of Pittsburgh, 3941 O'Hara Street, Pittsburgh, PA 15260 USA}

\begin{abstract}
Energy injection into the early universe can induce turbulent motions of
the primordial plasma, which in turn act as a source for gravitational
radiation. Earlier work computed the amplitude and characteristic
frequency of the relic gravitational wave background, as a function
of the total energy injected and the stirring scale of the turbulence.
This paper computes the frequency spectrum of relic gravitational
radiation from a turbulent source of the stationary Kolmogoroff form which
acts for a given duration, making no other approximations.
We also show that the limit of long
source wavelengths, commonly employed in aeroacoustic problems, is an
excellent approximation.
The gravitational waves from cosmological
turbulence around the electroweak energy scale
will be detectable by future space-based laser interferometers
for a substantial range of turbulence parameters.
\end{abstract}


\maketitle

\section{Introduction}

Direct detection of a relic gravitational wave  background is a
subject of considerable current interest (see \cite{m00,sources},
and \cite{Efstathiou06,hogan06,hughes06} for recent reviews),
motivated by planned satellite detection missions in the near
future \cite{LISA}. Gravitational wave detection could probe
directly the physical conditions in the early universe at the
epoch of radiation generation \cite{tasi}, since after been
generated, gravitational radiation freely propagates throughout
the entire evolution of the universe. Once generated, any
gravitational wave spectrum retains its shape, with all
wavelengths simply scaling with the expansion of the universe.
Various possibilities for early-universe physics leading to
detectable cosmological gravitational wave backgrounds include
quantum fluctuations during inflation \cite{inflation} and subsequent
oscillating classical fields during reheating \cite{reheating};
cosmological defects \cite{topological}; bubble wall motions and
collisions during phase transitions \cite{bubble,kos,kos2};
plasma turbulence \cite{kos2,kmk02,dolgov,cd06}; and cosmological
magnetic fields \cite{magnet,cd06}. Depending on wavelength, the
resulting gravitational waves might be detected either directly
or through their imprint on the polarization of the cosmic
microwave background \cite{kks-det}. If detected, gravitational
radiation generated in the early universe would provide a remarkable
new window into physics beyond the standard model of particle physics
(e.g., \cite{gs06,rs06}).

In this paper we revisit the generation of a cosmological
gravitational wave background from turbulent motion of the
primordial plasma. We employ methods similar to those originally
developed in aeroacoustics for calculating sound generation by
turbulent flows \cite{L52,P52,G,my75}. This allows us to
incorporate the influence of the temporal characteristics of
turbulent fluctuations on the gravitational wave generation
process, and thus to determine the spectrum
of the emitted gravitational waves at all frequencies. For
simplicity, we assume isotropic non-helical turbulence, ignoring
all possibilities for generating polarized gravitational waves
\cite{kgr05}. (Polarized radiation might be generated through
anisotropic stress of the helical primordial magnetic field
\cite{cdk04}, or from other parity-violating sources in the early
universe such as Chern-Simons coupling \cite{chern-simons,ay07,kam99}
or an axion field coupling with gravity \cite{aps04}. Detection
of these polarized backgrounds are discussed in Ref.~\cite{seto}.)

As is well known, gravitational waves are sourced by the
transverse and traceless part of the stress-energy tensor (see,
e.g., \cite{mtw73}). In our case
the stress-energy tensor results from turbulent plasma motions:
\begin{equation}
T_{ij}({\bf x}) \propto {\rm w} v_i({\bf x}) v_j({\bf x}),
\label{stresstensordef}
\end{equation}
where ${\bf v}({\bf x})$ is the velocity vector field of the
fluid and  ${\rm w}=p+\epsilon$ is the enthalpy density with $p$ and
$\epsilon$ the pressure and the energy density of plasma, which
is assumed to be constant throughout space \cite{kmk02}. To model
a period of cosmological turbulence, we assume that at time
$t_*$ in the early universe, a vacuum energy density
$\rho_{\rm{vac}}$ is converted into (turbulent) kinetic energy of
the cosmological plasma  via stirring on a characteristic source
length scale $L_S$,  over a time scale
$\tau_{\rm{stir}}$ \cite{kos}. The characteristic length scale
$L_S$ of the generated fluctuations is directly related to the
Hubble length $H^{-1}_*=H^{-1}(t_*)$ at the time of
energy injection. We consider only a forward cascade: after being
generated on the length scale $L_S$, the turbulence kinetic
energy cascades from larger to smaller scales. The cascade stops
at some damping scale $L_D$, when the energy of turbulence
thermalizes due to some dissipation mechanism, such as viscosity
or plasma resistivity. In this paper we consider ${\rm w}$,
$\rho_{\rm vac}$, $\tau_{\rm stir}$, $H_*$, $L_S$, and
$L_D$ as phenomenological parameters which can approximately
describe any period of cosmological turbulence, and derive the
dependence of the gravitational wave spectrum on these
parameters.  As expected from the universal nature of turbulence,
the shape of the spectrum scales with the characteristic
amplitude and frequency of the gravitational radiation.

We perform the computation of the gravitational wave spectrum in
real space, instead of using conventional Fourier space
techniques as in Ref.\ \cite{kmk02}. This makes the physical
interpretation of all quantities straightforward. The spatial
structure of the turbulence is taken to be isotropic with a
Kolmogoroff spectrum \cite{kol41}, and the time dependence of the
turbulence is described by the Kraichnan time auto-correlation
function \cite{K64}. While relativistic turbulence in the early
universe might depart somewhat from these scalings, these
assumptions are based on observed properties of laboratory
turbulence and will give the correct qualitative features of the
resulting radiation spectrum. Generalization to alternative
turbulence models is straightforward. We use natural units $\hbar
= c = k_B \equiv 1$ throughout.

\section{General Formalism}

We assume the duration of the turbulence, $\tau_T$, is much less than the Hubble time
$H^{-1}_\star$ \cite{kmk02,dolgov}, so the effects of the
expansion of the universe may be neglected in the generation of
gravitational radiation. This adiabatic assumption will be valid
for any turbulence which is produced in a realistic cosmological
phase transition \cite{tww91}. (Note that the duration of the turbulence $\tau_T$
can be substantially longer than the stirring time $t_{\rm stir}$; see the detailed
discussion in \cite{kmk02}.) Then the radiation equation in
real space can be written as \cite{mtw73}
\begin{equation}
\nabla^2 h_{ij}({\mathbf x}, t) -\frac{\partial^2}{\partial t^2}
h_{ij}({\mathbf x}, t) = -16\pi G S_{ij} ({\mathbf x}, t).
\label{eq:01}
\end{equation}
where $h_{ij}({\mathbf x}, t)$ is the tensor metric perturbation,
the traceless part of the stress-energy tensor $T_{ij} ({\mathbf
x}, t)$ is \cite{W}
\begin{equation}
S_{ij} ({\mathbf x}, t) = T_{ij} ({\mathbf x}, t) -\frac{1}{3}
\delta_{ij} T^k_k ({\mathbf x}, t), \label{eq:02}
\end{equation}
and $t$ is physical time. During the period of turbulence, the
stress tensor takes the form of Eq.~(\ref{stresstensordef}) \cite{kol41}.

The general solution of Eq. (\ref{eq:01}) is \cite{W,mtw73},
\begin{equation}
h_{ij}({\mathbf x}, t) = 4G \int {\rm d}^3 {\bf x}^\prime \frac{
S_{ij}({\mathbf x}^\prime,t-|{\bf x}- {\bf x}^\prime|) }{|{\bf x}-
{\bf x}^\prime|} . \label{hsol}
\end{equation}
Due to the stochastic character of the turbulent stress tensor
$S_{ij}$, the generated metric perturbations $h_{ij}$ also are
stochastic. We aim to derive the energy density spectrum of these
perturbations at the end of the turbulent phase; after that the
amplitude and wavelength of the gravitational radiation scales
simply with the expansion of the universe. The energy density of
gravitational waves is defined as \cite{m00}
\begin{equation}
\rho_{GW}({\bf x},t)= \frac{1}{32\pi G} \langle
\partial_t h_{ij}({\mathbf x},t) \partial_t h_{ij} ({\mathbf x},t)\rangle
= \frac{G}{2\pi } \int {\rm d}^3 {\bf x}^\prime {\rm
d}^3 {\bf x}^{\prime \prime}
\frac{\langle \partial_t S_{ij}({\mathbf x}^\prime,t^\prime) \partial_t S_{ij}({\mathbf
x}^{\prime \prime},t^{\prime \prime}) \rangle }
{|{\bf x}- {\bf x}^\prime| |{\bf x}- {\bf x}^{\prime \prime}|},
\label{eq:04}
\end{equation}
where the brackets denote an ensemble average over realizations of the
stochastic source, $t^\prime=t-|{\bf x}- {\bf x}^\prime|$
and $t^{\prime \prime}=t-|{\bf x}- {\bf x}^{\prime\prime}|$.

\subsection{Localized Source}

We will first consider turbulence in a bounded region of space
centered around ${\bf x}=0$. In this case,
the energy density flux ${\bf P}({\bf x},t)$ of the radiation
propagating outward in the direction $\hat {\bf n}$ is just
\begin{equation}
{\bf P}({\bf x}) = {\hat {\bf n}} \rho({\bf x},t). \label{eq:05}
\end{equation}
At large distances from the turbulent source, the far-field
approximation is justified \cite{W,mtw73}.
This assumption replaces $|{\bf x}-{\bf x}^\prime|$ by
$|\bf x|$ in Eq.~(\ref{eq:04}), yielding for the gravitational wave
energy density flux
\begin{equation}
{\bf P}({\bf x}) =\frac{G {\hat {\bf n}} }{2\pi |{\bf x}|^2} \int
{\rm d}^3 {\bf x}^\prime {\rm d}^3 {\bf x}^{\prime \prime} \langle
\partial_t S_{ij}({\mathbf x}^\prime,t^\prime)
\partial_t S_{ij}({\mathbf x}^{\prime \prime},t^{\prime \prime}) \rangle.
\label{eq:06}
\end{equation}
The flux from a spatially bounded source drops as the inverse
 square of the distance from the radiation source, as expected.

The autocorrelation function of the tensor metric perturbations is defined as
\begin{equation}
L({\bf x},\tau)\equiv \frac{1}{32\pi G} \langle
\partial_t h_{ij}({\mathbf x},t) \partial_t h_{ij} ({\mathbf x},t+\tau) \rangle,
\label{Ldef}
\end{equation}
with $\tau=t^\prime-t$,
such that $\rho_{GW} ({\bf x})=L({\bf x},0)$.
Defining the usual Fourier transform of $L({\bf x},\tau)$ as
\begin{equation}
I({\bf x},\omega) = \frac{1}{2\pi} \int {\rm d} \tau e^{i\omega
\tau} L({\bf x},\tau), \label{eq:08}
\end{equation}
with $\omega$ as the angular frequency, it readily follows that
\begin{equation}
\rho_{GW}({\bf x}) = \int {\rm d} \omega I({\bf x},\omega),
\label{eq:09}
\end{equation}
and therefore $I({\bf x},\omega)$ represents the spectral energy
density of induced gravitational waves \cite{mtw73,m00}.

Substituting Eq.~(\ref{hsol}) into Eq.~(\ref{Ldef}) gives
\begin{equation}
L({\bf x},\tau) =\frac{G }{2\pi |{\bf x}|^2} \int {\rm d}^3 {\bf
x}^\prime {\rm d}^3 {\bf x}^{\prime \prime} \langle
\partial_t S_{ij}({\mathbf x}^\prime, t^\prime)
\partial_t S_{ij}({\mathbf x}^{\prime \prime},t^{\prime \prime}) \rangle.
\label{eq:10}
\end{equation}
For the case of stationary turbulence, it can be proven that \cite{G}
\begin{equation}
\langle \partial_t S_{ij}({\mathbf x}^\prime, t^\prime)
\partial_t S_{ij}({\mathbf x}^{\prime \prime},t^{\prime \prime})
\rangle = - \partial_\tau^2 \langle S_{ij}({\mathbf x}^\prime,
t^\prime) S_{ij}({\mathbf x}^{\prime \prime},t^{\prime \prime})
\rangle. \label{eq:12}
\end{equation}
Using Eq.~(\ref{eq:12}) with the far-field approximation
 $|{\bf x}- {\bf x}^\prime| = |{\bf x}| - {\bf x} \cdot {{\bf x}^\prime}/|{\bf x}|$, and
using the fact  that the cross-correlation of a stationary random
function is independent of time translation, Eq.~(\ref{eq:10}) reduces to
\begin{equation}
L({\bf x},\tau) =\frac{-G}{2\pi |{\bf x}|^2} \partial_\tau^2 \int
{\rm d}^3 {\bf x}^\prime {\rm d}^3 {\bf x}^{\prime \prime} \langle
S_{ij}({\mathbf x}^\prime, t) S_{ij}({\mathbf x}^{\prime
\prime},\tau^\prime) \rangle, \label{eq:14}
\end{equation}
where
\begin{equation}
\tau^\prime = t+\tau + \frac{ {\bf x} }{|{\bf x}|} \cdot ( {{\bf
x}^{\prime \prime}} - {\bf x}^\prime). \label{eq:15}
\end{equation}

Defining the two-point time-delayed forth order correlation tensor
by
\begin{equation}
R_{ijkl}({\bf x}^\prime, {\bm \xi}, \tau) = \frac{1}{{\rm w}^2}
\langle S_{ij}({\bf x}^\prime ,t) S_{kl}({\bf x}^{\prime \prime},
t+\tau) \rangle, \label{eq:16}
\end{equation}
where ${\bm \xi}={\bf x}^{\prime \prime}-{\bf x}^\prime$
and ${\rm w}=\rho + p$ is the enthalpy density of the
plasma, Eq.~(\ref{eq:14}) yields
\begin{equation}
L({\bf x},\tau) =\frac{-G{\rm w}^2}{2\pi |{\bf x}|^2}
\partial_\tau^2 \int {\rm d}^3 {\bf x}^\prime {\rm d}^3 {\bf \xi}
R_{ijij} \left( {\bf x}^\prime, {\bm \xi}, \tau + \frac{{\bf
x}}{|{\bf x}|} \cdot {\bf \xi} \right). \label{eq:17}
\end{equation}
Fourier transforming this equation gives
\begin{equation}
I({\bf x},\omega) =\frac{4\pi^2\omega^2 G {\rm w}^2}{|{\bf x}|^2}
\int {\rm d}^3 {\bf x}^\prime H_{ijij} \left( {\bf x}^\prime,
\frac{{\bf x}}{|{\bf x}|} \omega, \omega\right) \label{eq:20}
\end{equation}
(summation on $i$ and $j$ assumed), where the four-dimensional
power spectral energy density tensor of stationary turbulence is
defined as
\begin{equation}
H_{ijkl}({\bf x}^\prime,{\bf k},\omega) \equiv \frac{1}{(2\pi)^4} \int
{\rm d}^3 {\bm \xi} {\rm d} \tau e^{i(\omega \tau - {\bf k}\cdot {\bm
\xi})} R_{ijkl}({\bf x}^\prime, {\bm\xi}, \tau).
\label{hijkldef}
\end{equation}
Equation (\ref{eq:20}) allows us to calculate the spectral
energy density of gravitational waves from a localized source,
if the real-space statistical properties of the turbulent
source are known.

\subsection{Spatially Homogeneous Source of Finite Duration}

For a cosmological source of stochastic gravitational radiation, we assume that
the source is statistically homogeneous, so that the averaged correlators of the
stress tensor have no spatial dependence, and isotropic, so that the correlator between
two spatial points depends only on the distance between the points and not on the
direction. We can also simply account for the
expansion of the universe by a simple rescaling of the frequency of all radiation
after its production, so we compute the radiation spectrum in a non-expanding spacetime
and include the expansion effect at the end.

With these assumptions, Eq.~(\ref{hijkldef}) simplifies to
\begin{eqnarray}
H_{ijkl}({\bf x}',{\bf k},\omega) &=& H_{ijkl}({\bf k},\omega)\nonumber\\
&=&\frac{1}{(2\pi)^4}\int d^3{\bm\xi}d\tau\,e^{i\omega\tau}e^{-i{\bf k}\cdot{\bm\xi}}R_{ijkl}({\bm\xi},\tau)
\label{hijklft}\\
&=&\frac{1}{4\pi^3}\int d\tau d\xi\,\xi^2 e^{i\omega\tau} j_0(k\xi) R_{ijkl}(\xi,\tau),
\label{hijklsimp}
\end{eqnarray}
so $H_{ijij}({\bf \hat x}\omega,\omega) = H_{ijij}(\omega,\omega)$ independent of the observation
direction $\bf\hat x$, as
expected on physical grounds. Now consider a stochastic source lasting for a finite duration
$\tau_T$, the duration of the turbulent source.
The total radiation energy spectrum at some point
and time is obtained by integrating over all source regions with a light-like separation from the observer,
which comprises a spherical shell around the observer with a thickness corrresponding to the duration of the phase transition, and a radius equal to the proper distance along any light-like path
from the observer to the source. Due to statistical isotropy and homogeneity, the integral
is trivial, contributing only a volume factor, giving for the total energy spectrum
\begin{equation}
\rho_{GW}(\omega) \equiv \frac{d\rho_{GW}}{d\,\ln\omega}=
16\pi^3\omega^3 G {\rm w}^2 \tau_T H_{ijij}(\omega,\omega).
\label{spectrumHijij}
\end{equation}
This spectrum is of course independent of the position of the observer, as it should
be for a stochastic background. In the absence of the expansion of the universe, a stochastic
source generates a spectrum of radiation which then remains constant for all later times.

\section{Statistics of Stationary Kolmogoroff Turbulence}

For a particular model of turbulent motion, the correlations
needed for computing gravitational radiation can be estimated.
Here we consider the simplest turbulence model, the original
Kolmogoroff picture. The spectral function $F_{ij}({\bf k},\tau)$
for stationary, isotropic and homogenous turbulence  is defined
as a spatial Fourier transform of the two-point velocity
correlation function
\begin{equation}
R_{ij}({\bf r},\tau) \equiv \langle v_i({\bf x}, t) v_j({\bf
x}+{\bf r}, t+\tau) \rangle.
\label{Rijdef}
\end{equation}
This function can be expressed in the form \cite{my75}
\begin{equation}
F_{ij}({\bf k},\tau)= \frac{E_k}{4\pi k^2}
\left(\delta_{ij}-\frac{k_i k_j}{k^2} \right) f(\eta_k,\tau),
\label{eq:4.1}
\end{equation}
where $E_k$ is the one-dimensional turbulent spectrum of energy
density,
$\eta_k$ is the autocorrelation function
\cite{K64}, and the function $f(\eta_k,\tau)$ characterizes
temporal decorrelation of turbulent fluctuations, such that it
becomes negligibly small for $\tau \gg 1/\eta_k$.

Here we consider Kolmogoroff turbulence for which the energy
density spectrum is given by the power law \cite{kol41}
\begin{equation}
E_k= C_K \varepsilon^{2/3} k^{-5/3},\qquad\qquad k_0<k<k_d,
\label{kol}
\end{equation}
defined over the range of wavenumbers from
$k_0$, determined by the stirring length scale $L_S\equiv 2\pi/k_0$
on which the energy is injected into turbulent motions,
to $k_d$, determined by the dissipation length scale $L_D\equiv 2\pi/k_d$
on which the plasma kinetic energy is thermalized. Here $C_K$
is a constant of order unity; for simplicity we set $C_K=1$. The
parameter $\varepsilon$ is the energy dissipation rate per
unit enthalpy, $\varepsilon\simeq \rho_{\rm vac}/(\tau_T {\rm w})$.
The corresponding autocorrelation function is \cite{my75}
\begin{equation}
\eta_k=\frac{1}{\sqrt{2\pi}} \varepsilon^{1/3} k^{2/3}.
\label{autocorrelation}
\end{equation}
We assume that the stirring and dissipation scales are well
separated,
i.e., $k_0 \ll k_d$, which corresponds to the turbulence having
high Reynolds number. This will be an excellent approximation
in any early universe phase transition with the stirring scale related to
the Hubble length.
For simplicity, we adopt Kraichnan's square exponential time dependence \cite{K64} to
model the temporal decorrelation,
\begin{equation}
f(\eta_k,\tau)=\exp \left( -\frac{\pi}{4} \eta^2_k \tau^2 \right).
\label{kraichnan}
\end{equation}
While other forms
of $f(\eta_k,\tau)$ are also frequently
used (see, e.g., \cite{Le}), neither total power of generated waves
nor the spectrum are very sensitive to the specific
form of the temporal decorrelation \cite{P52}.

To compute the fourth-order velocity correlation tensors Eq.~(\ref{eq:16}) needed
in the gravitational wave formula Eq.~(\ref{eq:20}), we invoke
the Millionshchikov quasi-normal hypothesis \cite{my75}:
\begin{equation}
\langle v_i^a v_j^a v_k^b v_l^b \rangle = \langle v_i^a v_j^a
\rangle \langle v_k^b v_l^b \rangle + \langle v_i^a v_k^b \rangle
\langle v_j^a v_l^b \rangle + \langle v_i^a v_l^b \rangle \langle
v_j^a v_k^b \rangle, \label{eq:4.4}
\end{equation}
where $v_i^a \equiv v_i({\bf x},t)$ and $v_i^b \equiv v_i({\bf x}+
{\bf r},t+\tau)$.
Using Eqs. (\ref{eq:02}), (\ref{eq:16}) and (\ref{eq:4.4}) we obtain
\begin{equation}
R_{ijij}({\bf x}^\prime, {\bf x}^\prime + {\bf r}, \tau) =
R_{ii}({\bf r}, \tau) R_{jj}({\bf r}, \tau)+\frac{1}{3} R_{ij}({\bf
r}, \tau) R_{ij}({\bf r}, \tau).
\label{Rijij_from_Rij}
\end{equation}
Then Eq.~(\ref{hijklft}) can be evaluated using
Eqs.~(\ref{eq:4.1})-(\ref{autocorrelation}) and the convolution
theorem to give
\begin{equation}
H_{ijij}({\bf k},\omega) = \frac{1}{6}\int {\rm d} {\bf k}_1
{\rm d} \omega_1 g({\bf k}_1,\omega_1) g({\bf k} - {\bf k}_1,\omega - \omega_1)
\left[ 27 - \frac{k^2}{k_1^2}+ \frac{k^4}{2k_1^2u^2} + \frac{k_1^2}{2u^2} - \frac{k^2}{u^2}
+ \frac{u^2}{2k_1^2} \right],
\label{eq:4.6}
\end{equation}
where we have defined $u\equiv |{\bf k}-{\bf k}_1|$ and
\begin{equation}
g({\bf k},\omega) \equiv \frac{E_k}{4\pi^2 k^2\eta_k}
\exp \left(- \frac{\omega^2}{\pi \eta^2_k} \right). \label{gdef}
\end{equation}
Choose the vector ${\bf\hat k}$ as the axis for spherical coordinates $(\theta_1,\phi_1)$
of the ${\bf k}_1$ integral. The azimuthal angular integral over $\phi_1$
is trivial. The dependence on the direction of ${\bf k}_1$ is clearly
only through ${\bf k}\cdot{\bf k}_1$, so $H_{ijij}({\bf k},\omega)
=H_{ijij}(k,\omega)$. The other angular integral can be simplified
by changing variables from $\theta_1$ to $u$, giving
\begin{eqnarray}
H_{ijij}(k,\omega) &=& \frac{\pi}{3}\int dk_1 d\omega_1 g(k_1,\omega_1)
\left(\frac{27k_1}{k} - \frac{k}{k_1}\right)
\int_{|k_1-k|}^{k_1+k} du\, u g(u,\omega - \omega_1)\nonumber\\
&& + \frac{\pi}{3}\int dk_1 d\omega_1 g(k_1,\omega_1)
\left(\frac{k^3}{2k_1} + \frac{k_1^3}{2k} - kk_1\right)
\int_{|k_1-k|}^{k_1+k} du\, \frac{1}{u} g(u,\omega - \omega_1)\nonumber\\
&& + \frac{\pi}{6}\int dk_1 d\omega_1 g(k_1,\omega_1)\frac{1}{kk_1}
\int_{|k_1-k|}^{k_1+k} du\, u^3 g(u,\omega - \omega_1).
\label{Hijij_triple}
\end{eqnarray}
We need to integrate this expression numerically. The $\omega_1$
integral can be done analytically in terms of the error function;
the entire expression is reduced to an integral over two
dimensionless quantities in Appendix A, Eq.~(\ref{Hijij_2}). The
result scales with the stirring scale $k_0$, and depends on the
Mach number
 ${\rm M} = (\varepsilon/k_0)^{1/3}$ of the turbulence. Its dependence on the
dissipation scale $k_d$ is through the Reynolds number ${\rm R} =
(k_d/k_0)^{4/3}$; as expected from physical considerations, the
radiated power is almost completely independent of ${\rm R}$.
Numerical results are displayed in the next Section.

\section{Relic Gravitational Waves}

The previous Section and the Appendix has given an analytic
expression for the gravitational wave energy spectrum resulting
from a period of turbulence lasting a time $\tau_T$, stirred on a
scale $k_0$, with Reynolds
number ${\rm R}$ and Mach number ${\rm M}$. The only significant
approximation made is that the turbulence is stationary and acts
as a source of gravitational waves for a finite time interval; the
error made through this idealization is discussed below. In order
to improve on this approximation, it would be necessary to create
a detailed numerical model of the turbulent source, including incorporating
an actual stirring mechanism, such as colliding bubbles in a phase
transition. We have also assumed that the expansion of the
universe can be ignored during the turbulence; this should be a
good approximation for any realistic early-universe phase
transition. The main effect of expansion would be only to damp
the total energy in the turbulence by a modest fraction, assuming
the turbulence does not last much longer than a Hubble time.

\subsection{The Spectrum at the Present Epoch}

To obtain the present spectrum, the gravitational waves generated by the
turbulent source must be propagated through the expanding universe
until today. The wavelengths
of the gravitational waves simply scale with the scale factor $a$ of the universe,
while their total energy density evolves like $a^{-4}$ and their amplitude decays like $a^{-1}$.
From $\rho_{GW}(\omega)$, Eq.~(\ref{spectrumHijij}), we can form
$\Omega_G(\omega) \equiv \rho_{GW}(\omega)/\rho_c$, with the critical density
$\rho_c = 3H_0^2/8\pi G$. Then, changing to  linear frequency $f=\omega/2\pi$, a characteristic strain amplitude is conventionally defined as
\begin{equation}
h_c(f) = 1.263 \times 10^{-18} \left(\frac{1\,{\rm Hz}}{f}\right)\left[h_0^2\Omega_G(f)\right]^{1/2}
\label{hcdef}
\end{equation}
where $h_0$ is the current Hubble parameter $H_0$ in units of 100
${\rm km}\,{\rm sec}^{-1}{\rm Mpc}^{-1}$. From the computed
$h_c(f)$ at the epoch of the turbulence, given by a scale factor
$a_*$, the factor by which the amplitude and the frequency are
reduced  is
\begin{equation}
\frac{a_*}{a_0} = 8.0\times 10^{-16}\left(\frac{100}{g_*}\right)^{1/3}\left(\frac{100\,{\rm GeV}}{T_*}\right),
\label{scalingfactor}
\end{equation}
where $T_*$ is the temperature of the universe with scale factor $a_*$, and $g_*$ is the
effective number of relativistic degrees of freedom the universe has at this time. To give
expressions which are physically transparent, we write the turbulence stirring scale and
the turbulence duration as fractions of the Hubble length during the turbulence:
\begin{equation}  \gamma H_*^{-1} = 2\pi/k_0,\qquad\qquad \zeta H_*^{-1} = \tau_T;
\label{gammadef}
\end{equation}
in other words, $\gamma$ is the stirring scale's fraction of the
Hubble length and $\zeta$ is the turbulence duration's fraction
of the Hubble length. For any particular angular frequency $\omega_*$ of the radiation
at the time of the
phase transition, we can then convert  $\omega_*$  and $h_c(\omega_*)$ to the amplitude
$h_c(f)$ and frequency $f$ of the relic gravitational wave
background today using the useful expressions for a
radiation-dominated universe
\begin{equation}
{\rm w} = \frac{4\rho_*}{3} = \frac{2\pi^2}{45}g_* T_*^4,\qquad\qquad
H_* = 1.66 g_*^{1/2}\frac{T_*^2}{m_{\rm Pl}}
\label{helpers}
\end{equation}
to get
\begin{equation}
f = 1.65\times 10^{-3}\,{\rm Hz}\,
\left(\frac{\omega_*}{k_0}\right)
\left(\frac{g_*}{100}\right)^{1/6}
\left(\frac{\gamma}{0.01}\right)^{-1} \left(\frac{T_*}{100\,{\rm
GeV}}\right), \label{ftoday}
\end{equation}
\begin{equation}
h_c(f) = 1.28\times 10^{-19} \left(\frac{100\,{\rm GeV}}{T_*}
\right) \left(\frac{100}{g_*}\right)^{1/3}
\left(\frac{\gamma}{0.01}\right)^{3/2}
\left(\frac{\zeta}{0.01}\right)^{1/2} \left[k_0^3 \omega_\star(f)
H_{ijij}(\omega_*(f), \omega_\star(f))\right]^{1/2}.
\label{hctoday}
\end{equation}

\begin{figure}
\includegraphics[width=6.5in]{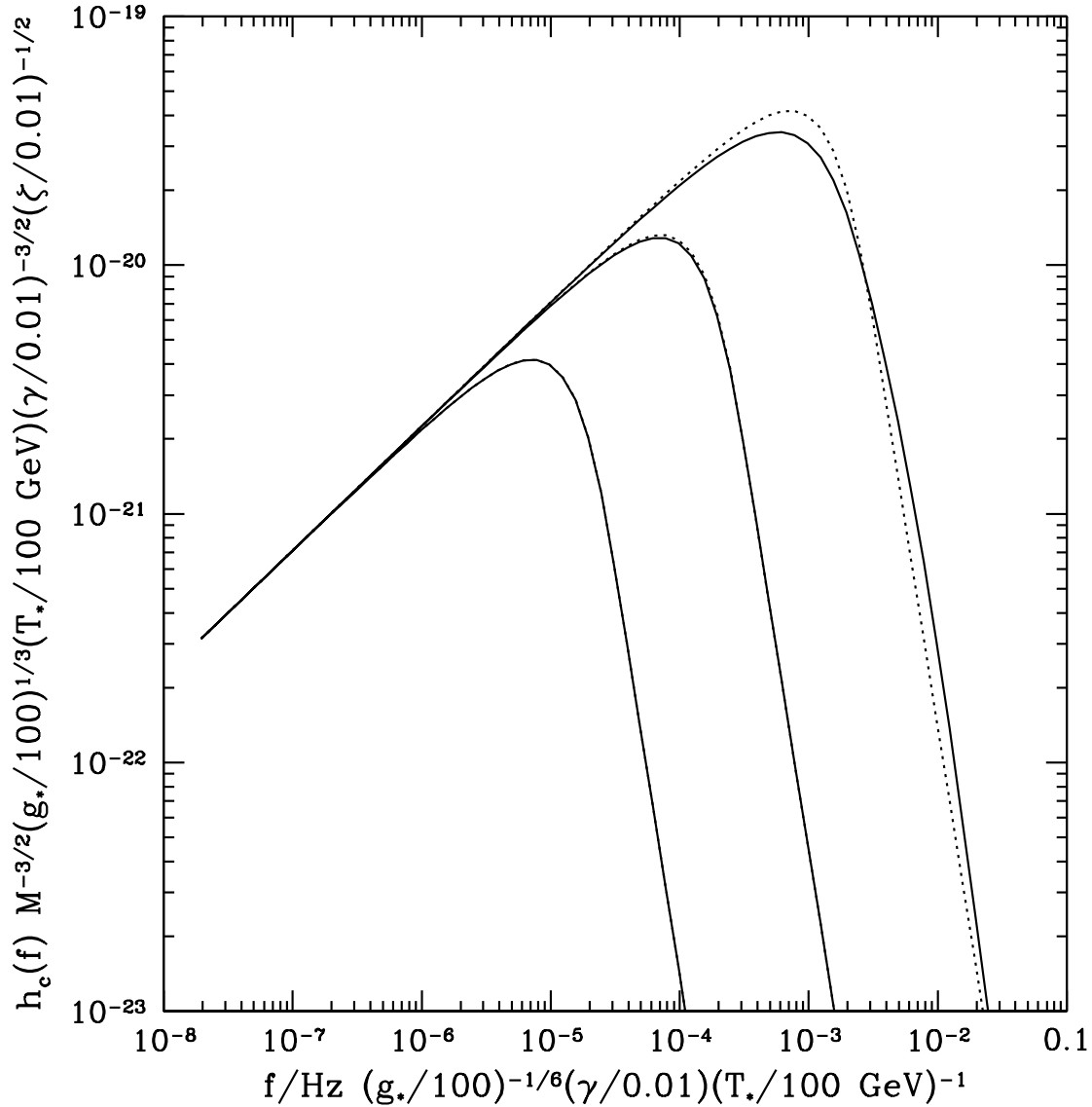}
\caption{The spectrum of gravitational radiation from turbulence.
The three solid lines are for different Mach numbers, with ${\rm M}=0.01$,
${\rm M}=0.1$, and ${\rm M}=1$ from lowest to highest amplitude. Note that these
three cases have also been scaled by a factor of ${\rm M}^{-3/2}$ for display, since
this is how the low-frequency tail scales with ${\rm M}$.
The dotted lines,
which are virtually indistinguishable from the solid lines except for
the ${\rm M}=1$ case, show the $k=0$ approximation to the gravitational
wave source.}
\label{fig:spectrum}
\end{figure}

The characteristic strain spectrum $h_c(f)$ is plotted in Fig.~1. The solid lines show
three different values for the Mach number, ${\rm M}=0.01$, ${\rm M}=0.1$, and ${\rm M}=1$, from lowest
to highest amplitude. This dependence on ${\rm M}$ is in addition to the explicit ${\rm M}^3$ scaling
in Eq.~(\ref{Hijij_2}), which is accounted for in the y-axis units.
The peak frequency of the spectrum scales inversely
with the stirring scale and linearly with the characteristic fluid velocity, which
is proportional to the Mach number. The peak frequency is thus proportional
to the inverse of the circulation time on the stirring scale of the turbulence.
This is the usual result for radiation generation: the characteristic frequency of radiation is
determined by the characteristic time scale of the source.

The characteristic parameter values to which the numbers in the plot
are scaled ($T_* = 100 \,{\rm GeV}$, $g_*=100$, $\gamma = \zeta = 0.01$) are values
consistent with turbulence arising from a strongly first-order phase transition at the
electroweak scale; see \cite{kos2} for a detailed discussion of the appropriate
parameters.

\subsection{The Aeroacoustic Limit}

Also plotted in Fig.~1 is an approximation common in
aeroacoustics \cite{G}, which replaces $H_{ijij}(k=\omega,\omega)$
with $H_{ijij}(k=0,\omega)$. It is clear that this simplifying
approximation is very good for ${\rm M}\leq 0.1$, and
overestimates the maximum amplitude  of $h_c(f)$ by around 30\%
for ${\rm M}=1$.
In this limit, the argument of the Bessel function in Eq.~(\ref{hijklsimp}) becomes small.
Substituting Eq.~(\ref{hijklsimp}) into Eq.~(\ref{spectrumHijij}) gives
\begin{equation}
\rho_{GW}(\omega) = 4\omega^3 G {\rm w}^2\tau_T \int d\tau d\xi\,
\xi^2e^{i\omega\tau} j_0(\omega\xi)R_{ijij}(\xi,\tau).
\label{energyspectrum}
\end{equation}
Thus if $\omega\xi$ is small compared to unity, the aeroacoustic limit $k\rightarrow 0$ is
guaranteed to be valid.

In the case of aeroacoustics, this approximation works because the
fluid velocity is always assumed to be small compared to the
velocity of the radiated acoustic waves (low Mach number). In the
cosmological regime, the interesting case is for plasma with a
relativistic amount of kinetic energy (otherwise there is not
substantial gravitational radiation produced). This will occur
only when the plasma is at a high enough temperature that it is
fully relativistic: otherwise, the amount of energy injected into
plasma motions would have to be a substantial fraction of the
particle mass scale rather than the cosmological temperature
scale, and this is unlikely on general grounds. A relativistic
plasma has sound speed $1/\sqrt{3}$, and the Mach number of the
turbulent plasma can never be much larger than 1; it will also
not be too much smaller than 1. In this case, the fluid
velocities will be roughly the sound speed, but this is close to
the propagation speed of the emitted radiation. Therefore, we do
not automatically have $\xi\omega\ll 1$ in
Eq.~(\ref{energyspectrum}), and the validity of the aeroacoustic
approximation must be ascertained by explicit calculation. As we
see in Fig.~1, the approximation still gives the right order of
magnitude for the spectrum amplitude even for Mach number ${\rm
M}=1$, corresponding to a fluid velocity equal to the sound speed.

\subsection{Asymptotic Limits}

The validity of the aeroacoustic approximation $k\rightarrow 0$ simplifies finding asymptotic forms for the
spectrum. Consider various frequency regimes of Eq.~(\ref{Hijij_triple}) with $k=0$; this is
facilitated by Eq.~(\ref{Hijijzero_2}). Note that in this limit, the dependence on the Mach
number ${\rm M}$ simply scales with the frequency. We assume ${\rm R} \gg 1$, or else fully developed turbulence cannot exist; this is an excellent approximation for early-universe plasma stirred on
scales near the Hubble length. In the low-frequency regime, simply take the limit $\omega\rightarrow 0$
to get
\begin{equation}
H_{ijij}(0,\omega)\sim \frac{7{\rm M}^3}{60k_0^4(\pi)^{3/2}},
\qquad\qquad \omega\rightarrow 0. \label{asymp_lowq}
\end{equation}
Physically, these frequencies are lower than the lowest characteristic frequency in the problem,
corresponding to the eddy turnover time on the stirring scale. This result of a constant $H_{ijij}$ is
universal and does not depend on either the spectrum or temporal characteristics of the turbulence
(see, e.g., Refs.~\cite{kos,kos2,kmk02,dolgov} and also \cite{cd06b}).
It translates to $h_c(f)$ scaling as $f^{1/2}$ at low $f$.

At high frequencies $\omega \gg k_0 {\rm M}{\rm R}^{1/2}$, the integral in Eq.(\ref{Hijijzero_2})
is dominated by the
contribution from its lower limit. After using the asymptotic
form ${\rm erfc}(x) \sim x^{-1}\pi^{-1/2}\exp(-x^2)$, $x\rightarrow\infty$,
an integration by parts gives the leading-order asymptotic behavior as
\begin{equation}
H_{ijij}(0,\omega) \sim \frac{7{\rm M}^3}{32\pi^2k_0^4{\rm
R}^{7/4}} \frac{k_0^2{\rm M}^2}{\omega^{2}}
\exp\left(-2\omega^2/(k_0^2{\rm M}^2{\rm R})\right), \qquad\qquad
\omega \gg k_0{\rm M}{\rm R}^{1/2}. \label{asymp_highq}
\end{equation}
This exponential suppression
is evident in Fig.~1; the dependence on ${\rm R}$ is negligible as expected from
physical considerations. The functional form of the high-frequency suppression is
determined by the specific form of the time autocorrelation function of the turbulence,
Eq.~(\ref{kraichnan}), but for any autocorrelation the amplitude of the emitted waves
should be very small in this regime. Physically, this limit corresponds to radiation frequencies
which are larger than any frequencies in the turbulent motions; consequently,
no scale of turbulent fluctuations generates these radiation frequencies directly, and the
resulting small radiation amplitude is due to the sum of small contributions
from many lower-frequency source modes. Since the integral is dominated by the
lower integration limit, the highest-frequency source fluctuations (which contain
very little of the total turbulent energy) contribute most to this high-frequency
radiation tail.

In the intermediate frequency regime, for frequencies
$k_0{\rm M} < \omega < k_0{\rm M}{\rm R}^{1/2}$,
the integral  in Eq.(\ref{Hijijzero_2}) is dominated by the contribution
around  $k_0^2{\rm M}^2/\omega^2$ due to the exponential factor
in the integrand, with a width of the same order.
Physically, this implies that radiation emission at some
frequency in this range is dominated by the turbulent vortices of the same frequency.
Consequently, we have the rough estimate
\begin{equation}
H_{ijij}(0,\omega)\simeq\frac{7{\rm M}^3}{16
k_0^4\pi^{3/2}}\left(\frac{k_0{\rm M}}{\omega}\right)^{15/2}.
\label{est_midq}
\end{equation}
This yields $h_c(f)\propto f^{-13/4}$, compared to $h_c(f)\propto
f_{\rm stir}^{-1/2}f^{-11/4}$ in Ref.~\cite{kmk02}, where $f_{\rm
stir}$ is the turbulence circulation frequency at the stirring
scale. The slight discrepancy from the spectrum shape in
Ref.~\cite{kmk02} comes about because we have
treated the time correlations of the turbulence in a more realistic
way.  Here we distinguish two time scales, the decorrelation time
which describes how the fluid velocities in a given size eddy are
correlated with each other after a given time interval, and the
largest eddy turnover time.
In practice,
the dropoff with frequency in this regime is strong enough that
the high-frequency behavior in Eq.~(\ref{asymp_highq}) only holds when
the spectrum is many orders of magnitude below the peak amplitude.

The intermediate frequency regime scaling with frequency depends
on the specific model of the turbulence power spectrum. The
Kolmogoroff model is not the only possibility, especially in the
presence of magnetic fields. Any model of turbulence which
includes the local transfer of energy in the wavenumber space
will satisfy $E_k^2/\eta_k\propto k^{-4}$ \cite{gog07}. In the
$k=0$ limit, it is straightforward to derive that in general,
$H_{ijij}(0,{\bar q})$ scales as $1/{\bar q}^{5/n}$, where $n$ is
the scaling exponent of the turbulence autocorrelation function.
For Kolmogoroff turbulence, $n=2/3$ (Eq.~(\ref{autocorrelation})).
But for Iroshnikov-Kraichnan turbulence (for example), $n=1$, and
consequently the frequency dependence is somewhat softer,
$H_{ijij}(0,{\bar q})\propto 1/{\bar q}^{5}$. In practical terms,
this modified turbulence spectrum produces radiation with
very similar detectability properties
to that from the Kolmogoroff turbulence spectrum considered here.

\section{Discussion}

We have calculated for the first time the spectrum of relic gravitational radiation resulting from
a period of stationary turbulence in the early universe, in terms of the turbulence duration,
stirring scale, Reynolds and Mach numbers, and the temperature of the universe
when the turbulence occurs. This is probably the best that can be done without a
detailed simulation of actual turbulent motions. The most likely source of
energy injection leading to turbulence is an early-universe phase transition; the
connection between the phenomenological parameters describing a phase
transition and the parameters describing the turbulence are given explicitly
in Ref.~\cite{kmk02}.

The calculation we present here is conceptually simple. The only assumptions
made are that the turbulence lasts for a finite duration which is at least a turnover time on the stirring scale, and that during this time the turbulence can be characterized as stationary.
The spatial power spectrum is taken to be the Kolmogoroff form, Eq.~(\ref{kol}),
with temporal correlations of the Kraichnan form, Eq.~(\ref{kraichnan}).
These scalings are appropriate for non-relativistic turbulence
with large Reynolds number. While the cosmological case will have large Reynolds
numbers, the turbulence will be relativistic in the most interesting cases for gravitational
radiation generation. As argued in Ref.~\cite{kmk02}, a non-relativistic approximation to relativistic turbulence likely underestimates the resulting radiation: relativistic turbulence contains more kinetic
energy for a given fluid velocity. We expect the same general results to hold, except
the expression for the Mach number ${\rm M}^3=\epsilon/k_0$ will clearly be modified,
giving smaller Mach numbers than this non-relativistic expression.

Cosmological turbulence will never
be precisely stationary, since the universe is expanding. Turbulence from a
phase transition will also not be stationary because the duration of the source
is comparable to the eddy turnover time on the stirring scale \cite{kmk02}, so the
turbulence will decay with time.
Even so, as long as the eddies on a given length scale can be treated
as uncorrelated sources of turbulence, Ref.~\cite{kmk02} argues that the resulting
radiation spectrum will be close to that from a stationary source, simply due to the
inevitable cascade of energy from the stirring scale down to the diffusion scale.
This point can be made somewhat more formally, using an argument similar to
that given by Proudman \cite{P52,my75}.
In the case of stationary turbulence, the time derivatives
in Eq.~(\ref{eq:06}) lead to factors of
$1/\tau_0$ when computing the radiation spectrum. If the
turbulence is decaying, then additional terms proportional to
time derivatives of the correlation functions also will appear.
But the characteristic time scale of the
turbulence decay $\tau_d$ is at least several times greater than the
turnover time on the stirring scale, and consequently, these additional terms
which are proportional to $1/\tau_d$ can be neglected compared to the
stationary term.

It has been claimed that turbulence in the early universe is a source of
such short duration that it cannot be treated as stationary at all, and that
the resulting radiation spectrum should be imprinted with the characteristic
wave number of the turbulent source instead of its characteristic frequency
\cite{cd06b}. We fully agree that a gravitational wave source lasting for
a sufficiently short duration will not be well described by a
short piece of a stationary source; clearly in the limit that the turbulence
duration $\tau$ goes to zero, the resulting radiation spectrum should
peak at a frequency corresponding to the characteristic wavenumber of
the source (the inverse stirring scale), and our formalism will not be
valid in that limit. However, it is straightforward to see that a turbulent
source in the early universe will actually last long enough so that our
calculations are valid. Ref.~\cite{cd06b} uses a simple toy model for a
stochastic cosmological gravity wave source to argue that the condition
$\tau_T\omega_s\approx 1$ represents the dividing line between sources
that imprint their characteristic frequency on the radiation and could be described
using our formalism, and sources that
imprint their wavenumber; here we write the duration of the source as $\tau_T$ and
the characteristic source angular frequency as $\omega_s$.
If we write the turbulence turnover time on the stirring scale as $t_{\rm stir}$, the
associated angular frequency is $\omega_s=2\pi/t_{\rm stir}$; also write the duration
of the turbulent source as $\tau_T = Nt_{\rm stir}$, so that it lasts $N$ times the stirring-scale
turnover time. We can thus write $\omega_s \tau_T = 2\pi N$, and even for the unrealistically
short duration $N=1$, $\omega_s\tau_T$ is significantly larger than unity. For realistic
turbulence, we expect the dissipation time to be multiple turnover times. Inspecting the
exact gravitational wave solution for the toy model source
in Ref.~\cite{cd06b} confirms that, even for $N$ as small as 1, the radiation spectrum will be
peaked at $\omega_s$.

We also assume that turbulence is non-magnetic and non-helical.
Either of these complications can modify the power law in
Eq.~(\ref{kol}) or the form of the time correlation
Eq.~(\ref{kraichnan}) \cite{kgr05}. The main effect of any
modification is to change the rate at which the radiation
spectrum falls off at high frequencies, but since this dependence
is quite steep, even substantial changes to the asymptotic
behavior of the spectrum lead to little qualitative difference in
the spectrum. As mentioned above, the low-frequency behavior is
independent of any details of the turbulence, and the peak
frequency is determined by the eddy turnover time on the stirring
scale where the energy density peaks, which will also be
independent of any details of the turbulent cascade.

The proposed Laser Interferometer Space Antenna (LISA) satellite mission has a 5$\sigma$
strain sensitivity to stochastic backgrounds of below $h_c=10^{-23}$ between frequencies
$10^{-3}$ and $10^{-2}$ Hz, and decreasing to around $h_c=10^{-20}$ at $10^{-4}$ Hz,
for one year of integration (see, e.g., \cite{lisasens}).
Comparing with Fig.~1, turbulence with a Mach number ${\rm M}=1$
would be a factor of 1000 larger than the LISA detection threshold at the peak frequency
around $10^{-3}$ Hz. For a Mach number ${\rm M}=0.1$,
the peak amplitude decreases by a factor of 100 due to the ${\rm M}^{-3/2}$ scaling and
the different signal spectrum. However, the peak frequency also shifts to $10^{-4}$ Hz,
at which point LISA's sensitivity has declined greatly; the steep high-frequency tail of
the gravitational wave spectrum makes detection with LISA marginal in this case. Detectors
consisting of two or more correlated LISA detectors or enhanced versions of LISA
optimized for detecting stochastic backgrounds have been discussed \cite{cl01}, such as the
envisioned GREAT mission \cite{great}; future space-based interferometers could be configured
to give strain sensitivities comparable to LISA, but with a frequency window between
$10^{-4}$ and $10^{-6}$ Hz. Such an experiment would easily detect turbulence at
the electroweak scale with a Mach number ${\rm M}=0.1$, and would even flirt with
a detection at ${\rm M}=0.01$. Turbulence generated at somewhat higher
energy scales shifts to higher frequencies and easier detection with LISA.

As is widely appreciated, detecting cosmological backgrounds of gravitational radiation is
not only an issue of detector sensitivity, but also of foreground discrimination. The
galactic population of short-period binaries of compact objects, mostly white dwarfs,
is known to produce a confusion-limited
stochastic background at frequencies below $10^{-3}$ Hz \cite{ben97}. At low frequencies,
separating this galactic source from a cosmological source is essential, likely by exploiting
the non-uniform directional distribution of an galactic source \cite{gia97,uv01,cor01}. A
uniform stochastic source arising from the confusion limit of numerous extragalactic binaries
provides a further complication \cite{sch01}, which can only be distinguished from a
primordial background via differing spectra. We also
note that the source of the turbulence itself may produce a gravitational wave spectrum, and that
the characteristic peak frequency may scale differently from the turbulent spectrum; see, e.g.,
the spectra for first-order phase transitions in Ref.~\cite{kos2}.
A distinctive two-peaked shape to the gravitational wave spectrum in certain
regions of parameter space will also aid in its detection.

We have no guarantees of violent events in the early universe. However, turbulence is
a completely generic result of energy injection on a characteristic length scale, and we
have shown in this paper that the resulting relic gravitational waves are within the realm
of detectability, even for turbulence with Mach numbers as low as 0.01, corresponding to
an energy input into the early universe of $10^{-4}$ of the total energy density. Many
scenarios for the electroweak phase transition \cite{apr02} and other physics \cite{rs06}
will result in releases of energy that are interestingly large. The remarkable possibility
of probing high-energy physics via the detection of vanishingly small spacetime
distortions left from when the universe was a trillionth of a second old impels us to look.

\appendix*

\section{Numerical Evaluation of $H_{ijij}(k,\omega)$}

We need to evaluate Eq.~(\ref{Hijij_triple}) explicitly, with $g(k,\omega)$ given
by Eq.~(\ref{gdef}). The integral over $\omega_1$ can be evaluated analytically
using the identity
\begin{equation}
\int_0^\infty dy\, \exp(-Ay^2) \exp(-B(x-y)^2)= \frac{1}{2}
\left(\frac{\pi}{A+B}\right)^{1/2}
\exp\left(-\frac{ABx^2}{A+B}\right) {\rm
erfc}\left(-\frac{Bx}{\sqrt{A+B}}\right). \label{omega1int}
\end{equation}
This expression is simple to derive by writing the integrand as a single
exponential and completing the square in the argument of the exponential,
followed by a linear change of variables to give the error function. Then
Eq.~(\ref{Hijij_triple}) becomes
\begin{eqnarray}
H_{ijij}(k,\omega)&=&\frac{\varepsilon}{24 \pi (2\pi)^{3/2}
k}\int_{k_0}^{k_d} dk_1 k_1^{-10/3}\int du\,u^{-10/3}
\left(k_1^{-4/3} + u^{-4/3}\right)^{-1/2} \left[27 -
\frac{k^2}{k_1^2} - \frac{k^2}{u^2} + \frac{k^4}{2k_1^2u^2}
+ \frac{k_1^2}{2u^2} + \frac{u^2}{2k_1^2}\right]\nonumber\\
&&\qquad\times\exp\left(-\frac{2\varepsilon^{-2/3}\omega^2}{k_1^{4/3}
+ u^{4/3}}\right) {\rm
erfc}\left(-\frac{2^{1/2}\varepsilon^{-1/3}\omega}{\left(k_1^{-4/3}+u^{-4/3}\right)^{1/2}}\right).
\label{Hijij_1}
\end{eqnarray}
The lower
limit on the $u$ integral is ${\rm max}[|k_1-k|,k_0]$ and the upper limit is
${\rm min}[k_1+k,k_d]$, provided the lower limit is less than the upper limit; otherwise
the integral over $u$ is zero. These conditions on the limits
arise due to the limited range of $k$ over which
the function $E_k$ has support.
Note that Eq.~(\ref{Hijij_1}) is regular as $k\rightarrow 0$, with the limit
\begin{equation}
H_{ijij}(0,\omega) =
\frac{7\varepsilon}{12\pi^{3/2}}\int_{k_0}^{k_d} dk_1 k_1^{-6}
\exp\left(-\frac{\omega^2}{\varepsilon^{2/3}k_1^{4/3}}\right)
{\rm erfc}\left(-\frac{\omega}{\varepsilon^{1/3}k_1^{2/3}}\right).
\label{Hijijzero_1}
\end{equation}

Now rescale all dimensionful quantities by powers of $k_0$ to make them
dimensionless; we abbreviate  $\varepsilon/k_0 = {\rm M}^3$, where M is
the Mach number of the turbulence, $k_d/k_0={\rm R}^{3/4}$, where R is the
Reynolds number of the turbulence, $p\equiv k/k_0$,
and $q=\omega/k_0$.
The change of variables $x=(k_1/k_0)^{-4/3}$, $y=(u/k_0)^{-4/3}$ simplifies
the remaining integrals, giving
\begin{eqnarray}
H_{ijij}(p,q) &=& \frac{3{\rm M}^3k_0^{-4}}{256(2\pi)^{3/2}p}
\int_{{\rm R}^{-1}}^1 dx x^{3/4}\int dy y^{3/4}(x+y)^{-1/2}
\exp\left(-\frac{2xy}{x+y}\frac{q^2}{{\rm M}^2}\right)
{\rm erfc}\left(-\frac{2^{1/2}y}{(x+y)^{1/2}}\frac{q}{\rm M}\right)\nonumber\\
&&\qquad\times\left[ 54 - 2p^2x^{3/2} - 2p^2y^{3/2} + p^4x^{3/2}y^{3/2}
+ \frac{x^{3/2}}{y^{3/2}} + \frac{y^{3/2}}{x^{3/2}}\right];
\label{Hijij_2}
\end{eqnarray}
the lower limit of the $y$ integral is ${\rm max}[(x^{-3/4}+p)^{-4/3},{\rm R}^{-1}]$
and the upper limit is ${\rm min}[|x^{-3/4}-p|^{-4/3},1]$, provided the lower limit is
less than the upper limit. In the limit $p\rightarrow 0$,
both of these limits are $x$, so the integral has a leading order behavior proportional to $p$
and thus $H_{ijij}(p,q)$ is regular, with the limit
\begin{equation}
H_{ijij}(0,q) \simeq \frac{7{\rm M}^3k_0^{-4}}{16
\pi^{3/2}}\int_{{\rm R}^{-1}}^1 dx\, x^{11/4} \exp\left(-{\bar
q}^2x\right) {\rm erfc}\left(-{\bar q}x^{1/2}\right),
\label{Hijijzero_2}
\end{equation}
where we have abbreviated ${\bar q}\equiv q/{\rm M}$ since in this limit the integral
depends only on ${\bar q}$ and not on either $q$ or ${\rm M}$ separately, aside from the
constant prefactor.  Note that ${\rm R}\gg 1$
for a medium which supports turbulence; we expect
${\rm R}>2000$ during the cosmological epochs of relevance. The integrals
converge as $R\rightarrow\infty$, and the lower limit of the x-integrals in
Eqs.~(\ref{Hijij_2}) and (\ref{Hijijzero_2}) can be replaced by zero.
In Eq.~(\ref{Hijij_2}), the terms
with factors of $x^{-3/4}$ and $y^{-3/4}$ in the integrand converge somewhat slowly but have small prefactors compared to the first term, giving a negligible
dependence of the integral on the diffusion scale.
Numerically, it is convenient to take $R$ as some large but finite value;
then the integrand in Eq.~(\ref{Hijij_2}) is smooth and regular over the full range of integration,
and can now be easily performed  for any values of $p$ and $q$.

\begin{acknowledgments}
We thank R. Durrer, G.~Gabadaze, A.~Gruzinov, D.~Grasso,
G.~Melikidze, and B.~Ratra for helpful discussions. G.G. and T.K.
acknowledge partial support from grant ST06/4-096 of the Georgian
National Science Foundation, and support from grant
061000017-9258 from The International Association for the
Promotion of Cooperation with Scientists from the Newly
Independent States of the Former Soviet Union (INTAS).
 A.K. gratefully acknowledges support
from NSF grant AST-0546035.
\end{acknowledgments}


\end{document}